\documentclass[11pt]{article}
\usepackage[english]{babel}
\usepackage{amsmath,amssymb,amsthm,graphicx}

\usepackage[width=\textwidth,indention=0mm,justification=centerlast,
 font=small,labelfont=bf,labelsep=period]{caption}

\usepackage[pdfpagemode=UseOutlines,
bookmarks=true, bookmarksopenlevel=3, bookmarksopen=true,
bookmarksnumbered=true, unicode=true,
pdffitwindow=true,pdfpagelayout=SinglePage,pdfhighlight=/P,
colorlinks=true,linkcolor=blue,citecolor=blue,urlcolor=blue]{hyperref}

\advance\textwidth 20mm  \advance\oddsidemargin -10mm
\advance\textheight 30mm \advance\topmargin -18mm

\allowdisplaybreaks[0]

\theoremstyle{plain}
\newtheorem{statement}{Statement}

\theoremstyle{remark}
\newtheorem{remark}{Remark}

\newcommand{\Real}{\mathbb{R}}
\newcommand{\Integer}{\mathbb{Z}}
\newcommand{\const}{\mathop{\rm const}}

\title{Some exact solutions of the Volterra lattice}

\author{V.E. Adler$^*$, A.B. Shabat\thanks{L.D. Landau Institute for Theoretical Physics, Chernogolovka, Russian Federation.\newline E-mail: adler@itp.ac.ru}}

\date{March 28, 2019}

\begin{document}\thispagestyle{empty}
\maketitle

\begin{abstract}
We study solutions of the Volterra lattice satisfying the stationary equation for its non-autonomous symmetry. It is shown that the dynamics in $t$ and $n$ are governed by the continuous and discrete Painlev\'e equations, respectively. The class of initial data leading to regular solutions is described. For the lattice on the half-line, these solutions are expressed in terms of the confluent hypergeometric function. The Hankel transform of the coefficients of the corresponding Taylor series is computed on the basis of the Wronskian representation of the solution.
\medskip

\medskip

\end{abstract}

%-------------------------------------------------------------------------------
\section{Introduction}

Let an evolution equation $u_t=f[u]$ admits a symmetry $u_\tau=g[t,u]$ (that is, vector fields $\partial_t$ and $\partial_\tau$ commute), then the stationary equation $g[t,u]=0$ defines a constraint which is consistent with the dynamics in $t$. Moreover, if $\partial_\tau$ belongs to a commutative Lie subalgebra of the higher symmetries, then the stationary equation inherits this subalgebra. It is well known that many important classes of exact solutions satisfy such stationary equations, including the finite-gap, multisoliton and rational solutions. On the other hand, if $\partial_\tau$ contains some members of the additional noncommutative Lie subalgebra of symmetries then the constraint leads to a Painlev\'e type equation. Such solutions are also considered in the literature for a long time, but, for the understandable reasons, they are studied much worse.

In this paper we study the solutions of the Volterra lattice $u_{n,t}=u_n(u_{n+1}-u_{n-1})$ which satisfy the stationary equation for the master-symmetry (plus lower order terms). An analogous, more simple constraint was studied in papers \cite{Its_Kitaev_Fokas_1990,Fokas_Its_Kitaev_1991}, where the evolution in $n$ was governed by the discrete Painlev\'e equation dP$_1$ and the evolution in $t$ was governed by the Painlev\'e equation P$_4$. In our case, the respective equations are dP$_{34}$ and P$_5$ (or P$_3$, for degenerating parameters). The corresponding set of solutions is not so small and contains a family of solutions depending on three essential parameters, which are regular for all $n,t$. 

Section \ref{s:constriants} contains the definition of the constraint under study, lowering of its order and reduction to the Painlev\'e equations. In Section \ref{s:regular}, we define a subclass of regular solutions. It is characterized by the special choice of initial data at the fixed singular point $t=0$ and by certain restrictions on the values of parameters, ensuring the absence of poles at $t\ne0$. These solutions describe small-scale oscillations in a region that grows linearly with increase of $t$. Such behavior is typical for generic solutions outside the soliton sector with the initial data in the form of sharp spikes. The constraint which we study is of interest as an example of exact solution (in terms of the Painlev\'e transcendents) for this mode. However, so far these solutions are investigated only numerically.

Section \ref{s:halfline} deals with the case when, in addition to the constraint equations, the condition $u_0=0$ is satisfied. Then the lattice equations admit a reduction on the half-line $n>0$, and the equations P$_5$ and P$_3$ are reduced, respectively, to the confluent hypergeometric equation and the Bessel equation. Section \ref{s:asymp} contains asymptotic formulas for the regular solutions related to the hypergeometric equation.

The class of solutions in study contains, in particular, the solution with the most simple initial data in the form of the unit step $u_0=0$, $u_n(0)=1$ for $n>0$. It was constructed in our previous paper \cite{Adler_Shabat_2018} by comparing of the Wronskian representation for the Volterra lattice solution \cite{Leznov_1980} with the well known result from the combinatorics about the Hankel transform for the Catalan numbers \cite{Aigner_1999, Stanley_1999, Layman_2001}. Now we are able to reverse this construction and to apply the constraint equations for computing of the Hankel transform. In Section \ref{s:det}, this is done for the coefficients of the Taylor expansion of the Kummer function.

%-------------------------------------------------------------------------------
\section{Constraints for the Volterra lattice}\label{s:constriants}

The Volterra lattice
\begin{equation}\label{ut}
 u_{n,t}=u_n(u_{n+1}-u_{n-1})
\end{equation}
possesses the symmetries
\begin{align*}
 u_{n,t_2}&=u_n(h_{n+1}-h_{n-1}),\quad h_n:=u_n(u_{n+1}+u_n+u_{n-1}),\\
 u_{n,\tau_0} &= tu_n(u_{n+1}-u_{n-1})+u_n,\\
 u_{n,\tau_1} &= tu_n(h_{n+1}-h_{n-1})
  +u_n\bigl(\bigl(n+\tfrac{3}{2}\bigr)u_{n+1}+u_n-\bigl(n-\tfrac{3}{2}\bigr)u_{n-1}\bigr).
\end{align*}
The flow $\partial_{\tau_0}$ corresponds to the scaling transformation and $\partial_{\tau_1}$ is the master-symmetry which generates the commutative Lie subalgebra of the symmetries by the formula $\partial_{t_{k+1}}=[\partial_{\tau_1},\partial_{t_k}]$, starting from $\partial_{t_1}=\partial_t$ \cite{Cherdantsev_Yamilov_1995, Adler_Shabat_Yamilov_2000}. Both sequences $\partial_{t_k}$ and $\partial_{\tau_k}$ are infinite (and the flows $\partial_{\tau_k}$ are non-local for $k>1$), but in this paper we will need only the above members of the whole hierarchy of the symmetries.

The stationary equation for any linear combination of the symmetries is a constraint compatible with (\ref{ut}). The equations which correspond to the commutative symmetries only, bring to the algebro-geometric (in particular, multisoliton) solutions. The simplest example involving a noncommutative symmetry is given by equation
\[
 u_{n,t_2}+2u_{n,\tau_0}=0
\]  
(the coefficient at the second term is fixed by scaling and the term $u_{n,t}$ can be neglected, due to the shift $t\to t-\const$). After dividing by $u_n$, the 5-point difference equation appears
\[
 h_{n+1}-h_{n-1} +2t(u_{n+1}-u_{n-1})+2=0.
\]
It can be easily reduced to the 3-point constraint
\begin{equation}\label{IKF}
 u_n(u_{n+1}+u_n+u_{n-1})+2tu_n+n+(-1)^nb+c=0,
\end{equation}
moreover, a straightforward computation proves that it is consistent with (\ref{ut}) if and only if the integration constants $b$ and $c$ do not depend on $t$. This equation, known as the discrete Painlev\'e equation dP$_1$, turns the lattice equations (\ref{ut}) into a coupled system for the variables $u_{n-1},u_n$ which is equivalent to the continuous Painlev\'e equation P$_4$ for the function $y=u_n$  \cite{Its_Kitaev_Fokas_1990, Fokas_Its_Kitaev_1991, Grammaticos_Ramani_1998, Grammaticos_Ramani_2014}:
\begin{equation}\label{P4}
 y''=\frac{(y')^2}{2y}+\frac{3}{2}y^3+4ty^2+2(t^2-\alpha)y+\frac{\beta}{2y},
\end{equation}
\[
 \alpha=\frac{1}{2}(n-3(-1)^nb+c),\quad \beta=-(n+(-1)^nb+c)^2.
\]
The mapping $(u_{n-1},u_n)\mapsto(u_n,u_{n+1})$ defines one of the B\"acklund transformations for (\ref{P4}).

In this paper, our main goal will be to investigate another, more complicated case
\[
 u_{n,\tau_1} -4au_{n,\tau_0} -du_{n,t}=0.
\]
Here, the coefficient $a$ can be scaled either to 0 or to 1, and the shift of $t$ makes possible to remove the term $u_{n,t_2}$. Like in the previous example, this 5-point constraint can be reduced to a 3-point one, although it is less obvious in this case. First, we notice that the equation takes the following form, after dividing by $u_n$:
\[
 \widetilde G_{n+1}+\widetilde G_n=0 \quad\Rightarrow\quad G_n=\widetilde G_n+4(-1)^nb=0, 
\]
where we denote
\begin{equation}\label{Gn}
 G_n=(q_{n+2}+q_{n+1})u_{n+1}-(q_n+q_{n-1})u_n-4a(q_{n+1}-q_n)+4(-1)^nb=0
\end{equation}
and
\begin{equation}\label{qn}
 q_n=2tu_n+n-d.
\end{equation}
Next, we lower the order by use of the integrating factor:
\[
 (q_{n+1}+q_n)G_n=\widetilde F_{n+1}-\widetilde F_n=0 \quad\Rightarrow\quad 
 F_n=\widetilde F_n-4c=0,
\]
where $F_n$ is the left hand side of equation (\ref{unn}) below.  

\begin{statement}
The lattice (\ref{ut}) is consistent with the equation
\begin{equation}\label{unn}
 F_n=(q_{n+1}+q_n)(q_n+q_{n-1})u_n-4(aq^2_n+(-1)^nbq_n+c)=0,
\end{equation}
where $q_n=2tu_n+n-d$, for any constants $a,b,c,d$ and under assumption that $q_{n+1}+q_n\ne0$ at some inital moment $t=t^*$, for all $n$.
\end{statement}
\begin{proof}
A straightforward computation proves the identities
\begin{align*}
 G_{n,t}&=u_{n+1}(G_{n+1}+G_n)-u_n(G_n+G_{n-1}),\\ 
 F_{n,t}&= u_n(q_n+q_{n-1})G_n+u_n(q_{n+1}+q_n)G_{n-1},
\end{align*}
assuming, like for the constraint (\ref{IKF}), that the integration constants $b$ and $c$ do not depend on $t$. Let equation $F_n=0$ is satisfied for $t=t^*$, then also $G_n=(F_{n+1}-F_n)/(q_{n+1}+q_n)=0$ for $t=t^*$. Then it follows from the above identities that $G_n=F_n=0$ for all $t$ such that the solution $u_n(t)$ exists.
\end{proof}

\begin{remark}\label{rem:confinement} 
The stipulation that $q_{n+1}+q_n$ do not vanish at $t=t^*$ is not very essential. We need it only in order to provide $G_n(t^*)=0$. However, in the situation when $q_{n+1}+q_n=0$ for some $n$, we can directly require that the initial conditions satisfy the equality $G_n=0$, then the constraint (\ref{unn}) will be preserved as before. In other words, if we consider equation (\ref{unn}) as a mapping $(u_n,u_{n+1})\mapsto(u_{n+1},u_{n+2})$ then the equality $q_{n+1}+q_n=0$ determines a singularity which is eliminated by use of the equation $G_n=0$.
\end{remark}

Similar to the case (\ref{IKF}), the constraint (\ref{unn}) turns the lattice equations (\ref{ut}) into a  coupled system for the variables $u_{n-1},u_n$, and the shift $n\mapsto n+1$ defines a B\"acklund transformation for the latter. This system is equivalent, after some additional changes, to the Painlev\'e equations P$_5$ or P$_3$, depending on the value of the parameter $a$. The following statement is verified by straightforward substitution, for which it is convenient to completely rewrite equations (\ref{ut}) and (\ref{unn}) it terms of $q_n$:
\begin{gather}
\label{qt}
 q_{n,t}=\frac{1}{2t}(q_n-n+d)(q_{n+1}-q_{n-1}),\\
\label{qnn}
 (q_{n+1}+q_n)(q_n+q_{n-1})= \frac{8t(aq^2_n+(-1)^nbq_n+c)}{q_n-n+d}.
\end{gather}
Notice, that (\ref{qnn}) coincides, up to a scaling of parameters, with dP$_{34}$ equation \cite{Grammaticos_Ramani_2014}. 

\begin{statement}
Let functions $q_n(t)$ satisfy equations (\ref{qt}), (\ref{qnn}). If $a\ne0$ then functions 
\[
 y_n(t)=1-\frac{8at}{q_{n+1}(t)+q_n(t)}
\]
satisfy the P$_5$ equation  
\begin{equation}\label{P5}
 y''=\Bigl(\frac{1}{2y}+\frac{1}{y-1}\Bigr)(y')^2-\frac{y'}{t}
    +\frac{(y-1)^2}{t^2}\Bigl(\alpha y+\frac{\beta}{y}\Bigr)
    +\gamma\frac{y}{t}+\delta\frac{y(y+1)}{y-1},
\end{equation}
with parameters
\[
 \alpha=\frac{b^2-4ac}{8a^2},\quad \beta=-\frac{(a+(-1)^nb)^2}{8a^2},\quad
 \gamma=-2a(2n-2d+1),\quad \delta=-8a^2.
\]
If $a=0$ then functions
\[
 y_n(z)=\frac{1}{2z}(q_{n+1}(t)+q_n(t)),\quad t=z^2,
\]
satisfy the P$_3$ equation: 
\begin{equation}\label{P3}
 y''=\frac{(y')^2}{y}-\frac{y'}{z} +\frac{1}{z}(\alpha y^2+\beta) +\gamma y^3+\frac{\delta}{y},
\end{equation}
\[
 \alpha=-4n+4d-2,\quad \beta=-4(-1)^nb-8c,\quad \gamma=4,\quad \delta=-16b^2.
\]
\end{statement}

%-------------------------------------------------------------------------------
\section{Regular solutions}\label{s:regular}

In general, solutions of equations (\ref{ut}), (\ref{unn}) may have singularities in $t$. Such solutions are of interest as well, but in this paper we restrict ourselves by consideration of a special family of solutions, such that functions $u_n(t)$ are continuously differentiable on the whole axis $t\in\Real$, for all $n\in\Integer$. This regularuty condition strictly fixes the initial data of the lattice at $t=0$, that is, at the singular point of the system (\ref{ut}), (\ref{unn}). Indeed, since the values $u_n(0)$ are finite for a regular solution, hence $q_{n+1}(0)+q_n(0)=2n-2d+1$ and we obtain from (\ref{unn}) that, if $d$ is not a half-integer then
\begin{equation}\label{und0}
 u_n(0)= a + \frac{4(-1)^nb(n-d)+4c+a}{4(n-d)^2-1},\quad d\not\in\frac{1}{2}+\Integer.
\end{equation}
Therefore, for the fixed values of $a,b,c$ and $d$, we are talking about one special solution of equations (\ref{ut}) and (\ref{unn}) (while the general solution is 2-parametric). In terms of the P$_5$ equation (\ref{P5}), this solution corresponds to the functions $y_n(t)$ without singularity at $t=0$, with the initial data
\[
 y_n(0)=1,\quad y'_n(0)=-\frac{2\delta}{\gamma}=-\frac{8a}{2n-2d+1}.
\]
In the case when $d=\frac{1}{2}+k$ is half-integer, the integrating factor $q_{n+1}+q_n$ which we used for derivation of (\ref{unn}) turns into 0 for $t=0$ and $n=k$, and we have to use equation (\ref{Gn}) instead of (\ref{unn}), as it was explained in Remark \ref{rem:confinement}. For $t=0$, it takes the form
\[ 
 (n-k+1)u_{n+1}(0)-(n-k-1)u_n(0)=2a-2(-1)^nb.
\]
From here, all $u_n(0)$ are uniqueley defined, except for $u_k(0)$ and $u_{k+1}(0)$ which are chosen independently on the rest values in such a way that their sum is constant:
\begin{equation}\label{unk0}
\begin{gathered} 
 u_n(0)=a+b\frac{(-1)^n(2n-2k-1)+(-1)^k}{2(n-k)(n-k-1)},~~ n\ne k,k+1,\\
 u_{k+1}(0)+u_k(0)=2a-2(-1)^kb,\quad d=\frac{1}{2}+k,~ k\in\Integer.
\end{gathered} 
\end{equation}
These inital data can be viewed as a limiting case of (\ref{und0}) with $c=(-1)^kb/2-a/4$, for $d\to\frac{1}{2}+k$.

Under certain relations between parameters, it is possible that the equality $u_m(0)=0$ is fulfilled which splits the lattice (\ref{ut}) into two independent systems for $n<m$ and $n>m$. We consider this case in the rest sections in more details, and now we will assume that $u_n(0)\ne0$ for all integer $n$.

The absence of singularity at $t=0$ does not guarantee that the solution is regular for all $t$, this requires an additional study. The numeric experiments show that, for the solution family under scrutiny, the crucial property is related with the signs of $u_n(0)$: is all $u_n(0)$ are of one and the same sign (positive, without loss of generality) then the solution is regular for all $n,t$; in contrast, if there are $u_n(0)$ with different signs then the solution acquires a singularity at a finte $t$.

\begin{remark}
The question about the regularity criterium fot the solutions of lattice equations (\ref{ut}) with generic inital data is open. Regular solutions with different signs do exist: a simple explicit example is the solution
\begin{equation}\label{expsol}
 u_{2n} = -\frac{\beta(n+\delta)e^{2\beta t}}{\alpha+e^{2\beta t}},\quad 
 u_{2n+1}=\frac{\beta(\alpha n+\gamma)}{\alpha+e^{2\beta t}},
\end{equation}
which, apparently, does not have singularities for $\alpha\ge0$, and also the stationary solution $u_{2n}=\alpha$, $u_{2n+1}=\beta$ with constants of different signs. However, the nonalternating solutions are of primary interest. In many papers, this requirement is simply postulated; sometimes, the Volterra lattice is wrtten down in the variables $p_n=\sqrt{u_n}$.
\end{remark}

\begin{figure}[t]
\centerline{\includegraphics[width=70mm]{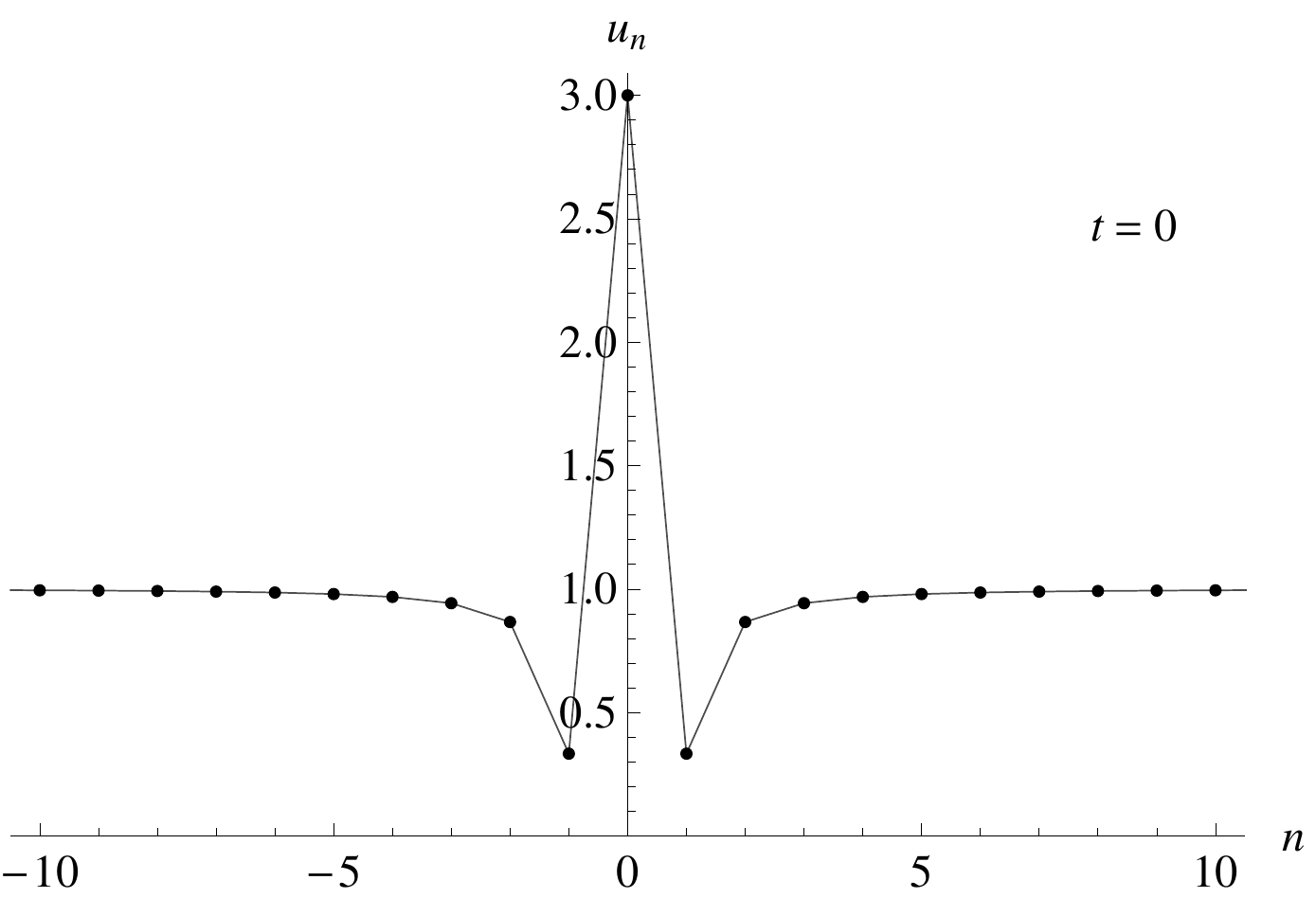}\quad\includegraphics[width=70mm]{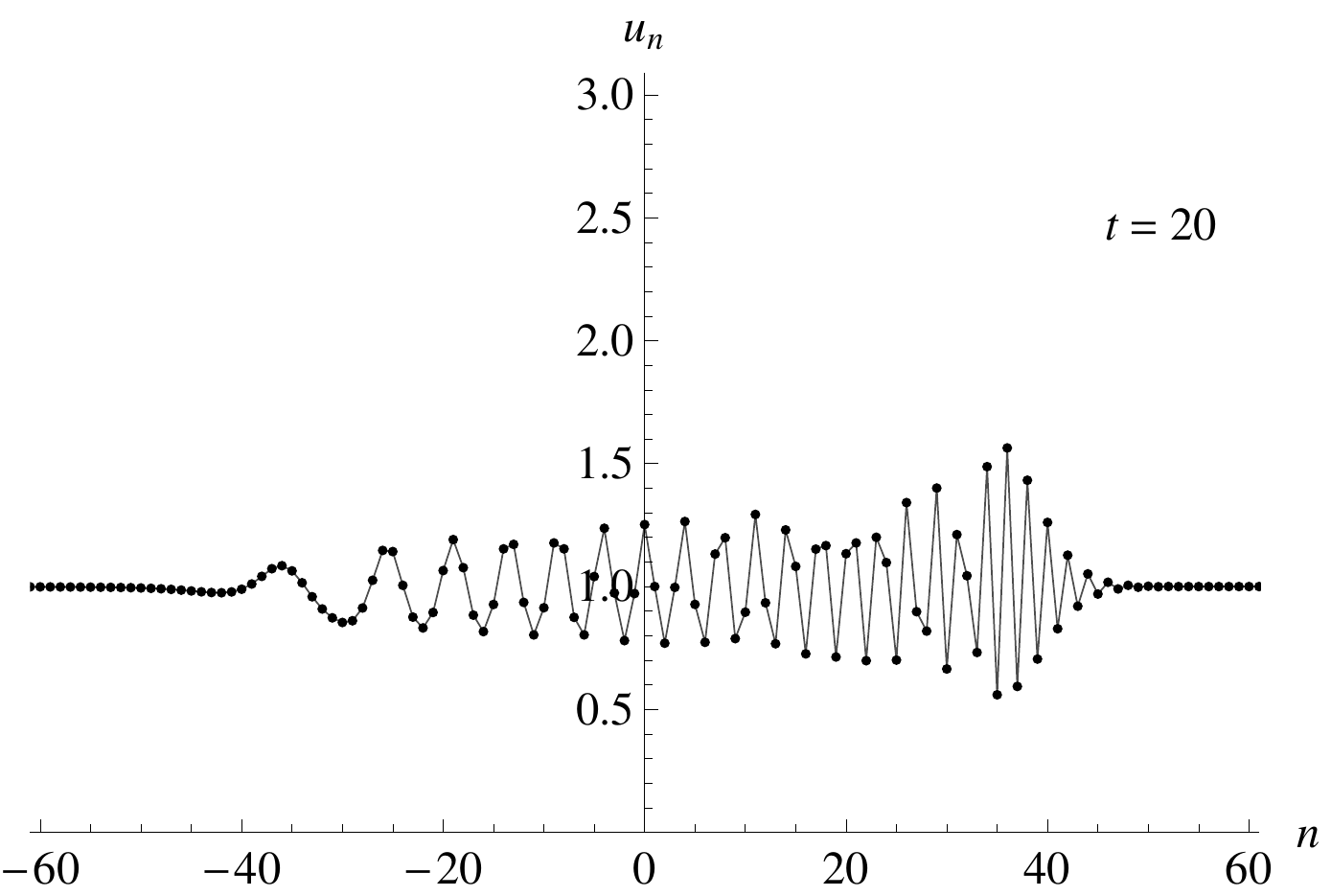}}
\caption{Solution of the Volterra lattice with the initial data $u_n(0)=1-\frac{2}{4n^2-1}$}
\label{fig:arrow}
\end{figure}

\begin{figure}[t]
\centerline{\includegraphics[width=70mm]{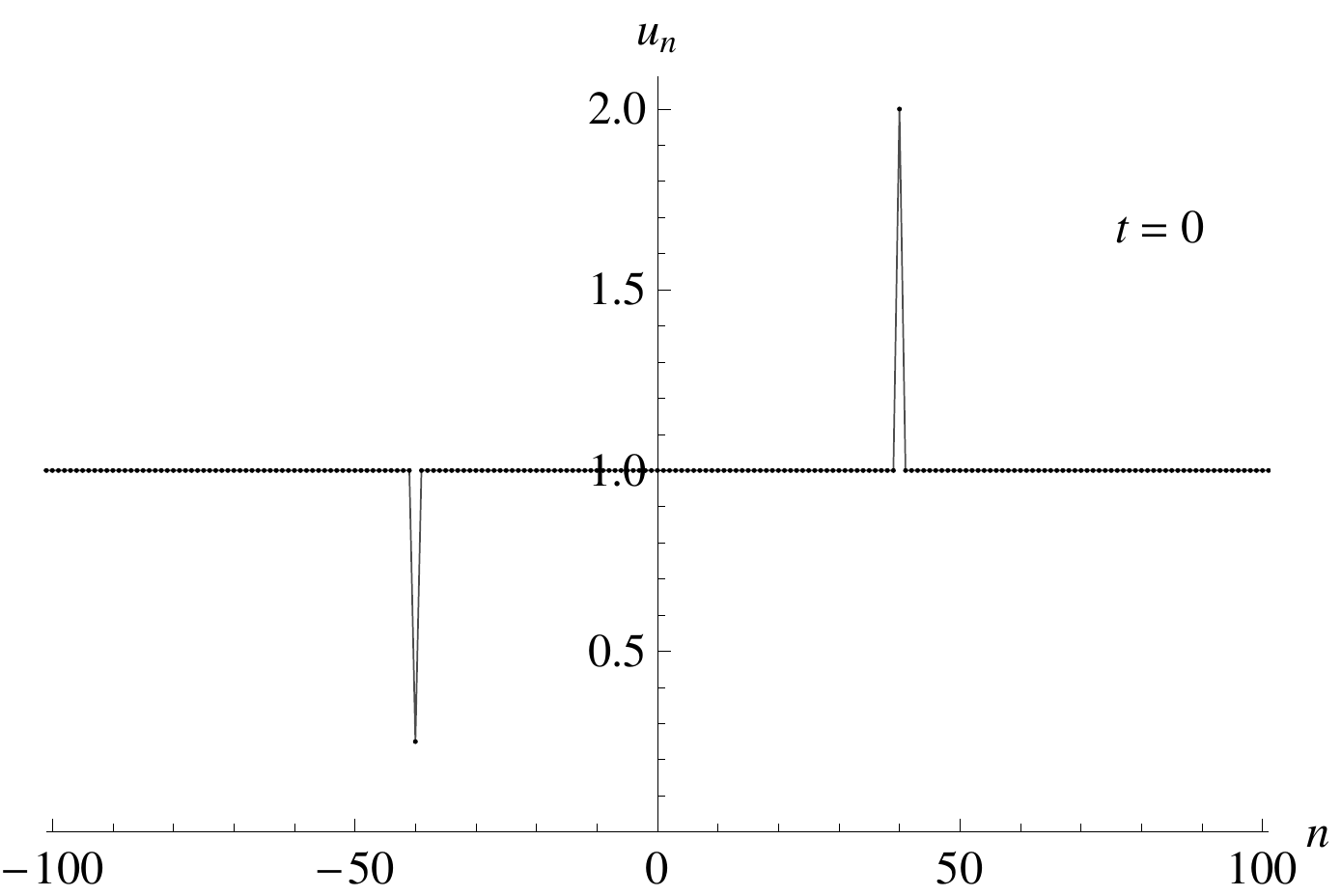}\quad\includegraphics[width=70mm]{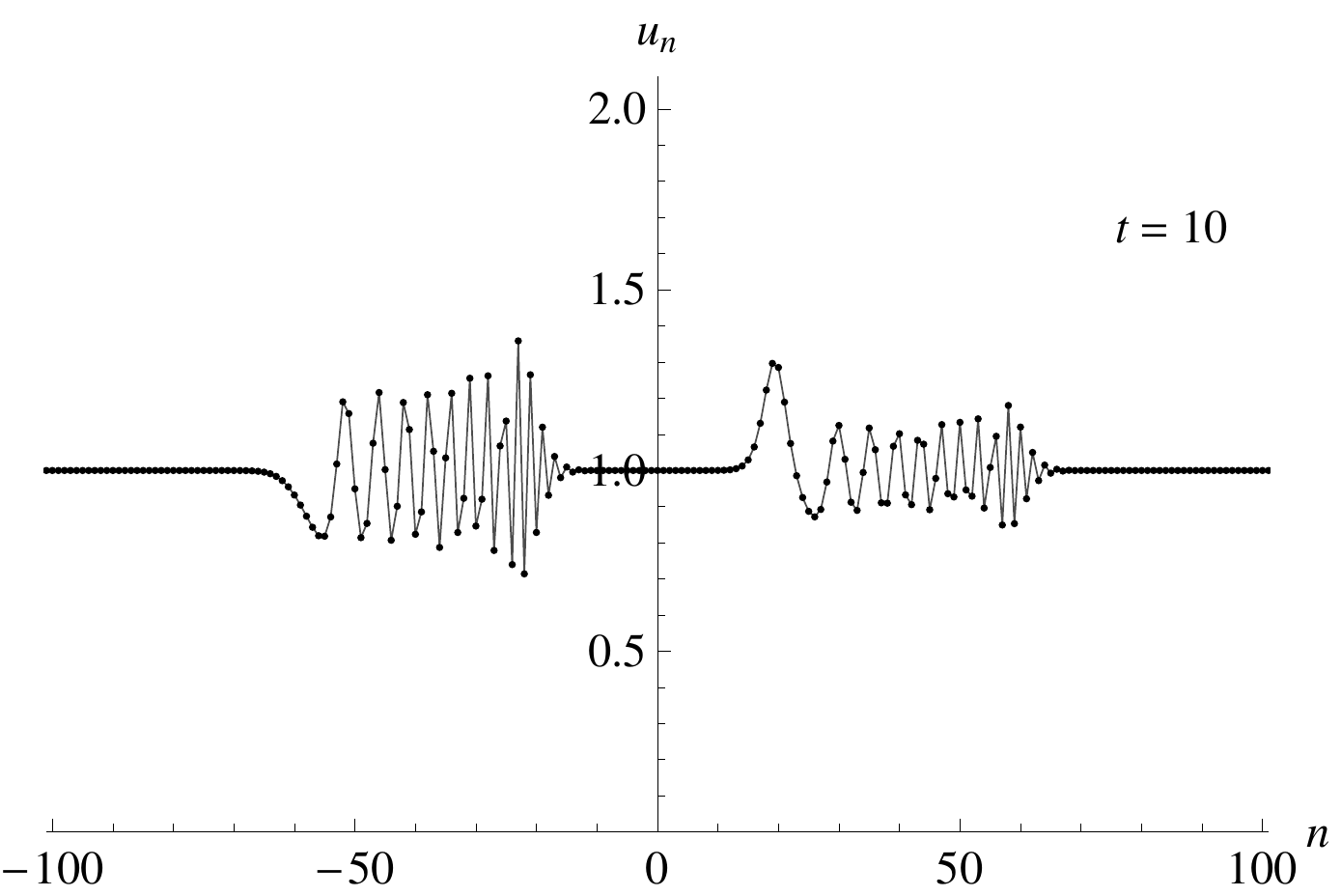}}\bigskip

\centerline{\includegraphics[width=70mm]{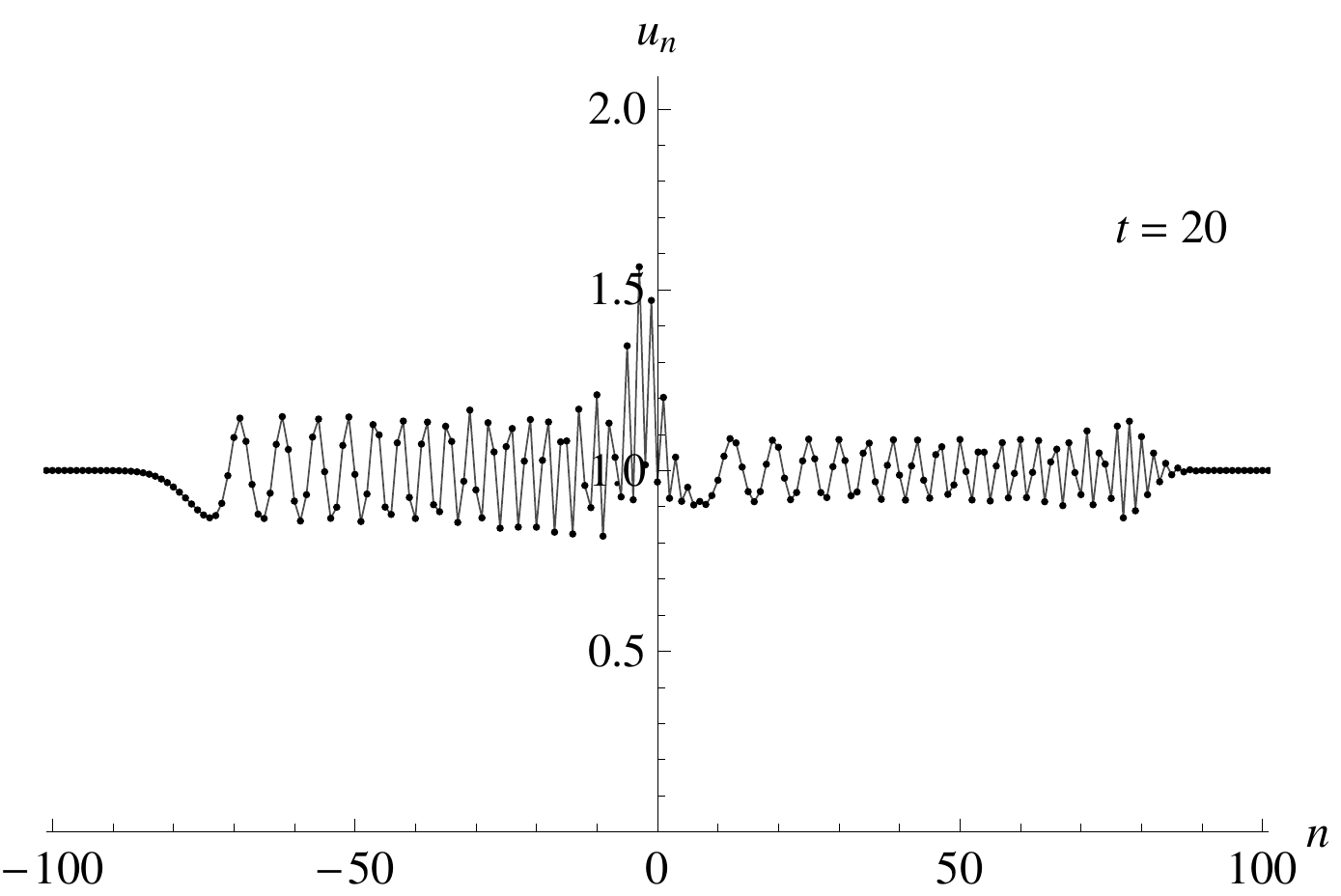}\quad\includegraphics[width=70mm]{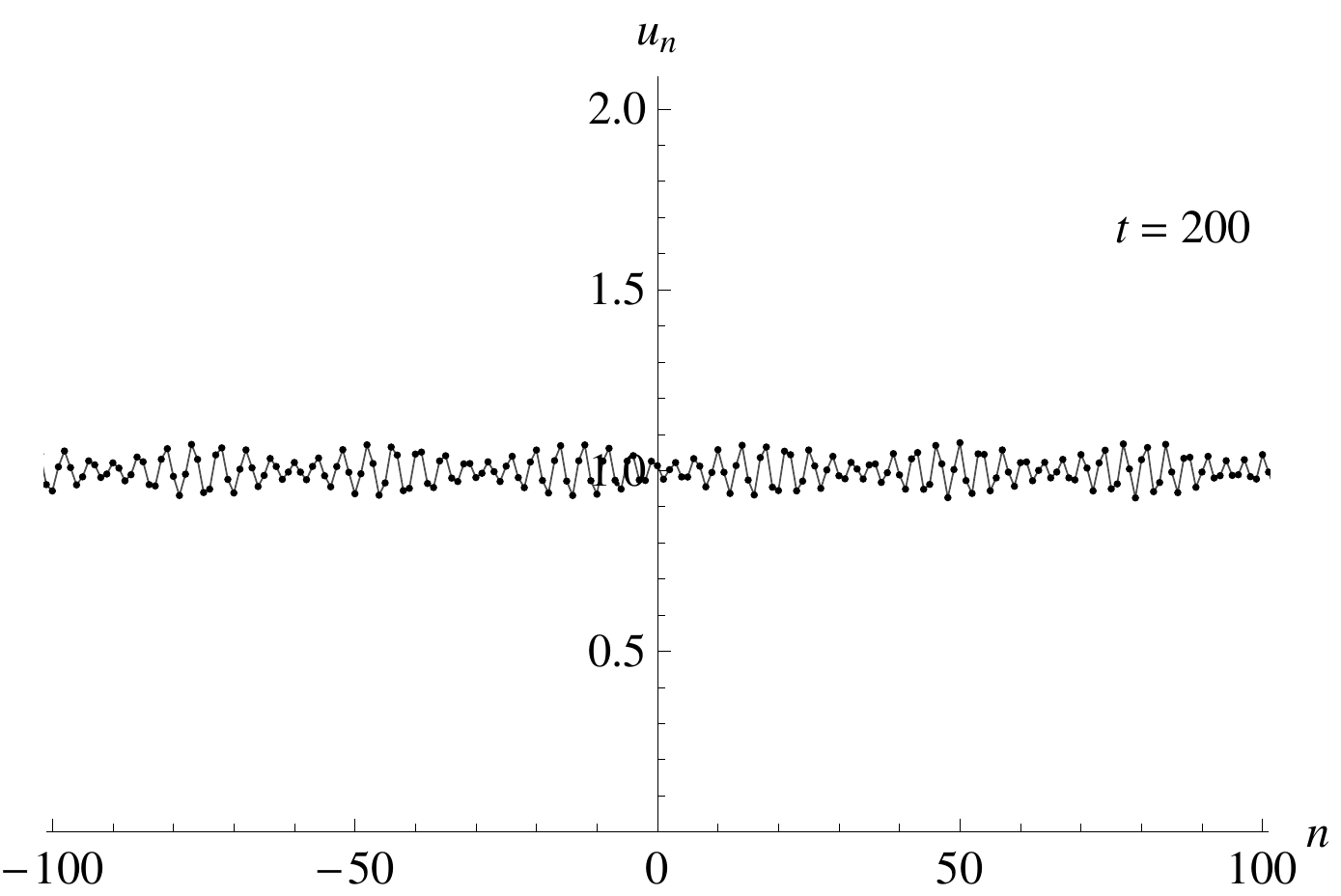}}
\caption{Solution of the Volterra lattice with the initial data\\ $u_{-40}(0)=0.25$, $u_{40}(0)=2$, $u_n(0)=1$, $n\ne\pm40$.}
\label{fig:2arrow}
\end{figure}

For $a=0$ and nonzero $b,c$, the initial data (\ref{und0}), (\ref{unk0}) always change the sign and there are no regular solutions. If $a\ne0$ then one can set $a=1$ without loss of generality. In addition, taking into account the shift of $n$, one can assume that $d\in(-\frac{1}{2},\frac{1}{2}]$. Then the condition of the positivity of the initial data reduces to inequalities
\[
 bd-c-d^2>0,\quad b(d-1)+c+(d-1)^2>0,\quad b(d+1)+c+(d+1)^2>0
\]
which cut off a bounded region in the parameter space (for a fixed $d$ it is a triangle in the $b,c$ plane) and the corresponding solutions are regular. Fig.\,\ref{fig:arrow} demonstrates a typical solution from this family, corresponding to the values $a=1$, $b=0$, $c=-3/4$ and $d=0$. The initial profile is shown on the left plot. It quickly collapses and generates a zone of small-scale oscillations (with the period comparable to the lattice spacing), which has an arrow-shaped profile and expands at a constant speed in both directions with increasing $t$ (for the negative $t$, the direction of the arrowhead changes). For nonzero $b$ and $d$, the initial data look a bit more complicated, but the general behavior of the solution remains the same. Moreover, the picture does not change much if we take the initial data that do not satisfy the constraint (\ref{unn}), but are close to (\ref{und0}). Apparently, this behavior is typical for solutions with generic initial data in the form of sharp spikes (as opposed to solutions of the soliton type, which are formed when the initial data are relatively gently sloping). Fig.\,\ref{fig:2arrow} shows a solution for initial data which differ from 1 at two points. Each spike generates oscillations of the type described, which form an interference pattern after fusion.

Thus, this is a fairly common mode in the Volterra lattice that deserves to be studied. It would be interesting to obtain its description from the point of view of the inverse scattering method. The constraint (\ref{unn}), with positive initial data, is of interest as an exact solution example for this mode, in terms of the Painlev\'e transcendents. More precisely, here we use not all solutions of P$_5$, but only one, which is distinguished by the regularity condition at $t=0$. However, this solution does not seem to be expressed in terms of classical special functions, at least for the general values of parameters.

%-------------------------------------------------------------------------------
\section{Solutions on the half-line}\label{s:halfline}

Assume that, in addition to the constraint (\ref{unn}), the condition $u_0=0$ is fulfilled (which is also a constraint, consistent with the lattice equations (\ref{ut})). In this case, the lattice splits into two unrelated subsystems for $n<0$ and $n>0$. It is enough to consider solutions on the half-line $n>0$. First of all, we notice that if $u_0=0$ then the system of ordinary differential equation for the variables $u_0,u_1$ (which is equivalent, in general, to a Painlev\'e equation) reduces to the Riccati equation for $u_1$.

\begin{statement}\label{st:Riccati}
Let $u_n$ be a solution of the lattice (\ref{ut}) governed by the constraints (\ref{unn}) and $u_0=0$. Then the function $u_1(t)$ satisfies the equation
\begin{equation}\label{u1'}
 u'_1+u^2_1-\Bigl(4a+\frac{2d-3}{2t}\Bigr)u_1-\frac{2(a-b)}{t}=0
\end{equation}
which is linearizable via the substitution $u_1=f'/f$:
\begin{equation}\label{f''}
 tf''+\Bigl(\frac{3}{2}-d-4at\Bigr)f'-2(a-b)f=0.
\end{equation}
\end{statement}
\begin{proof}
Substituting of $u_0=0$ into equations (\ref{unn}) for $n=0$ and $n=1$ gives
\begin{gather*}
 -ad^2+bd-c=0,\\
 \frac{1}{4}(q_2+q_1)(q_1-d)u_1 -aq^2_1 +bq_1 -c =0,
\end{gather*}
where $q_n=2tu_n+n-d$. Subtracting one equation from another and dividing by $q_1-d$, we obtain
\[
 \frac{1}{4}(q_2+q_1)u_1-a(q_1+d)+b=0.
\]
This is equivalent to (\ref{u1'}), taking into account the relation $u'_1=u_1u_2$ which follows from equation (\ref{ut}). The passage to equation (\ref{f''}) is standard.
\end{proof}

Thus, $u_0=0$ and $u_1$ is constructed by solving equation (\ref{f''}), then all functions $u_{n+1}$ for $n=1,2,\dots$ are constructed by recurrent relations, either by use of the lattice itself:
\begin{equation}\label{recD}
 u_{n+1}=\frac{u'_n}{u_n}+u_{n-1},
\end{equation}
or, without using differentation, by relation
\begin{multline}\label{rec0}
\quad u_{n+1}=-u_n-\frac{1}{t}(n-d+1/2)\\
   +\frac{a(2tu_n+n-d)^2+(-1)^nb(2tu_n+n-d)+bd-ad^2}{tu_n(tu_n+tu_{n-1}+n-d-1/2)}\qquad
\end{multline}
(this is the constraint equation (\ref{unn}) with $c=bd-ad^2$ resolved with respect to $u_{n+1}$). 

As in the previous section, we will consider only regular solutions, and such that $u_n\ne0$ for $n>0$ (since otherwise the solution is constrained to a finite interval). In this case, equations (\ref{und0}), (\ref{unk0}) are slightly refined and we arrive to the following statement. As before, it guarantees only regularity at $t=0$; the regularity for all $t$ is related with the constant sign property of the sequence (\ref{und00}).

\begin{statement}\label{st:un00}
Let $u_0=0$ and let $u_n\ne0$ be a solution of the lattice (\ref{ut}) for $n>0$, satisfying the constraint (\ref{unn}) and regular at $t=0$. Then
\begin{equation}\label{und00}
 u_1(0)=\frac{4(b-a)}{2d-3},~~ u_n(0)=\frac{a(n-d)^2+(-1)^nb(n-d)+d(b-ad)}{(n-d)^2-1/4},~~ n>1,
\end{equation}
where
\begin{equation}\label{bdnek}
 d\ne\frac{1}{2}+k,\quad b\ne a(2k-1),\quad b\ne 2a(d-k),\quad k=1,2,3,\dots
\end{equation}
\end{statement}
\begin{proof}
The expression for $u_n(0)$, $n>1$, is obtained from (\ref{und0}) by substituting $c=bd-ad^2$. Here, the values $d=\frac{1}{2},-\frac{1}{2},-\frac{3}{2},\dots$ are admissible, since the corresponding denominators do not vanish. If $d\ne\frac{1}{2}$ then the expression for $u_1(0)$ is also found from (\ref{und0}) and if $d=\frac{1}{2}$ then it is obtained from (\ref{unk0}) for $k=0$ (or directly from equation (\ref{u1'})). Solving the inequalities $u_{2k-1}(0)\ne0$ for the obtained initial data brings to restrictions $b\ne a(2k-1)$; the inequalities $u_{2k}(0)\ne0$ yield $b\ne 2a(d-k)$.

For $d=\frac{1}{2}+k$, $k>0$, we use equations (\ref{unk0}) instead of (\ref{und0}). In this case, if follows from the equality $u_0=0$ that
\[
 a=b\frac{2k+1-(-1)^k}{2k(k+1)}
\]
and it is easy to check that then also $u_{2k+1}(0)=0$, so these values of $d$ are rejected.
\end{proof}

If $a=0$ and $b\ne0$ then the change $z=2\sqrt{2bt}$, $f(t)=t^{d/2-1/4}y(z)$ brings (\ref{f''}) to the Bessel equation
\begin{equation}\label{Bessel}
 z^2y''+zy'+\bigl(z^2-(d-\tfrac{1}{2})^2\bigr)y=0.
\end{equation}
The corresponding initial data (\ref{und00}) are alternating. Numeric experiments show that if $b<0$ (without loss of generality) then the solution acquires the poles at $t<0$, but it tends to 0 for $t>0$ (the corresponding function $y$ is the Bessel function of the imaginary argument). This gives an example of alternating solution which is bounded and regular in the quadrant $n,t>0$. However, this solution is very unstable with respect to the calculation errors and the perturbations of the initial data.

%-------------------------------------------------------------------------------
\section{Asymptotics in the case $a=1$, $u_0=0$}\label{s:asymp}

If $a\ne0$ then the scaling of the independent variable $z=4at$, $f(t)=y(z)$ brings (\ref{f''}) to the confluent hypergeometric equation
\begin{equation}\label{confl}
 zy''+(\beta-z)y'-\alpha y=0,\quad \alpha=\frac{a-b}{2a},\quad \beta=\frac{3}{2}-d.
\end{equation}
Since the function $u_1=f'/f$ must be regular at $t=0$, we should to choose as $y$ the Kummer function $M(\alpha,\beta,z)$ (or ${\!}_1F_1(\alpha;\beta;z)$ in other notation), that is, $f(t)= M(\alpha,\beta,4at)$. 

\begin{figure}[t!]
\centerline{\includegraphics[width=70mm]{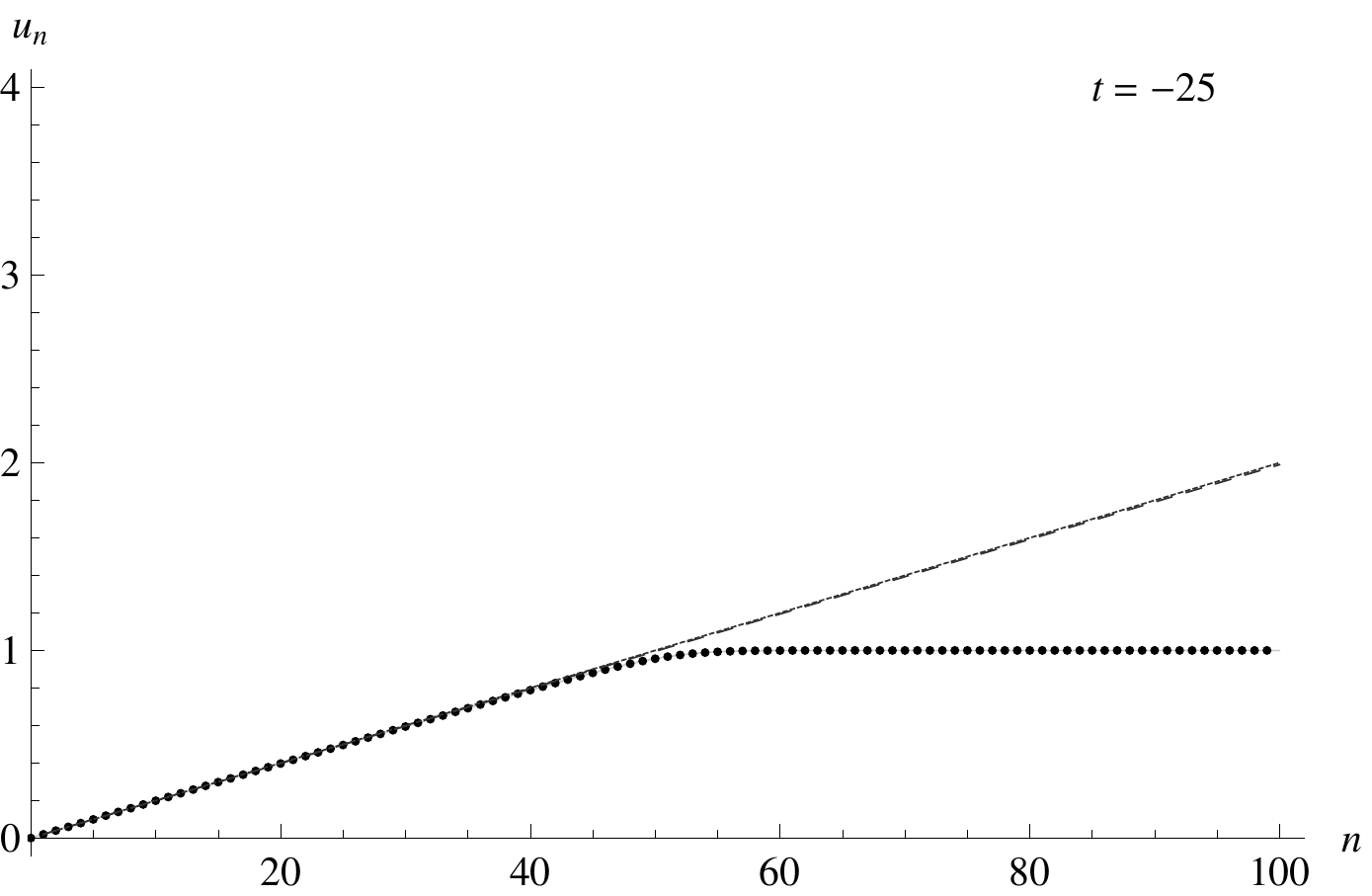}\qquad\includegraphics[width=70mm]{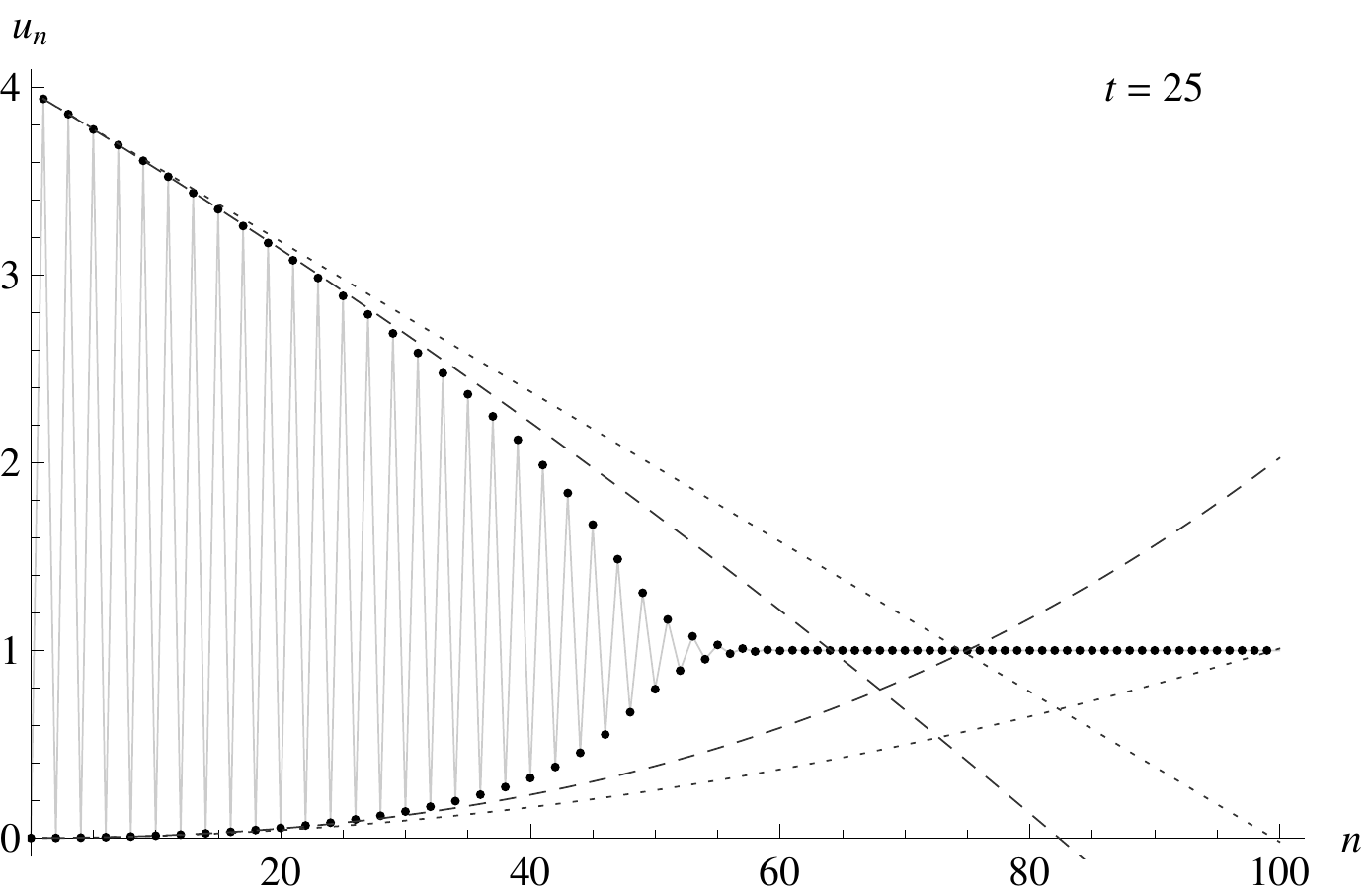}}
\caption{Solution with the initial data $u_n(0)=1$, $n>0$. The dashed lines correspond to one or two terms of the asymptotic expansions.}
\label{fig:t25}
\end{figure}

In this section we set $a=1$ without loss of generality. If $b=0$ and $d=-\frac{1}{2}$ then the initial data take especially simple form of the unit step: $u_n(0)=1$ for $n>0$. The corresponding solution, shown of fig.\,\ref{fig:t25}, was studied in \cite{Adler_Shabat_2018} (notice, that we used there an equivalent representation of $f$ in terms of the modified Bessel function: $f(t)=e^{2t}(I_0(2t)-I'_0(2t))=M(\frac{1}{2},2,4t)$; however, $f$ is not expressed through $I_n$ for the generic initial data (\ref{und00})). 

\begin{figure}[t!]
\centerline{\includegraphics[width=70mm]{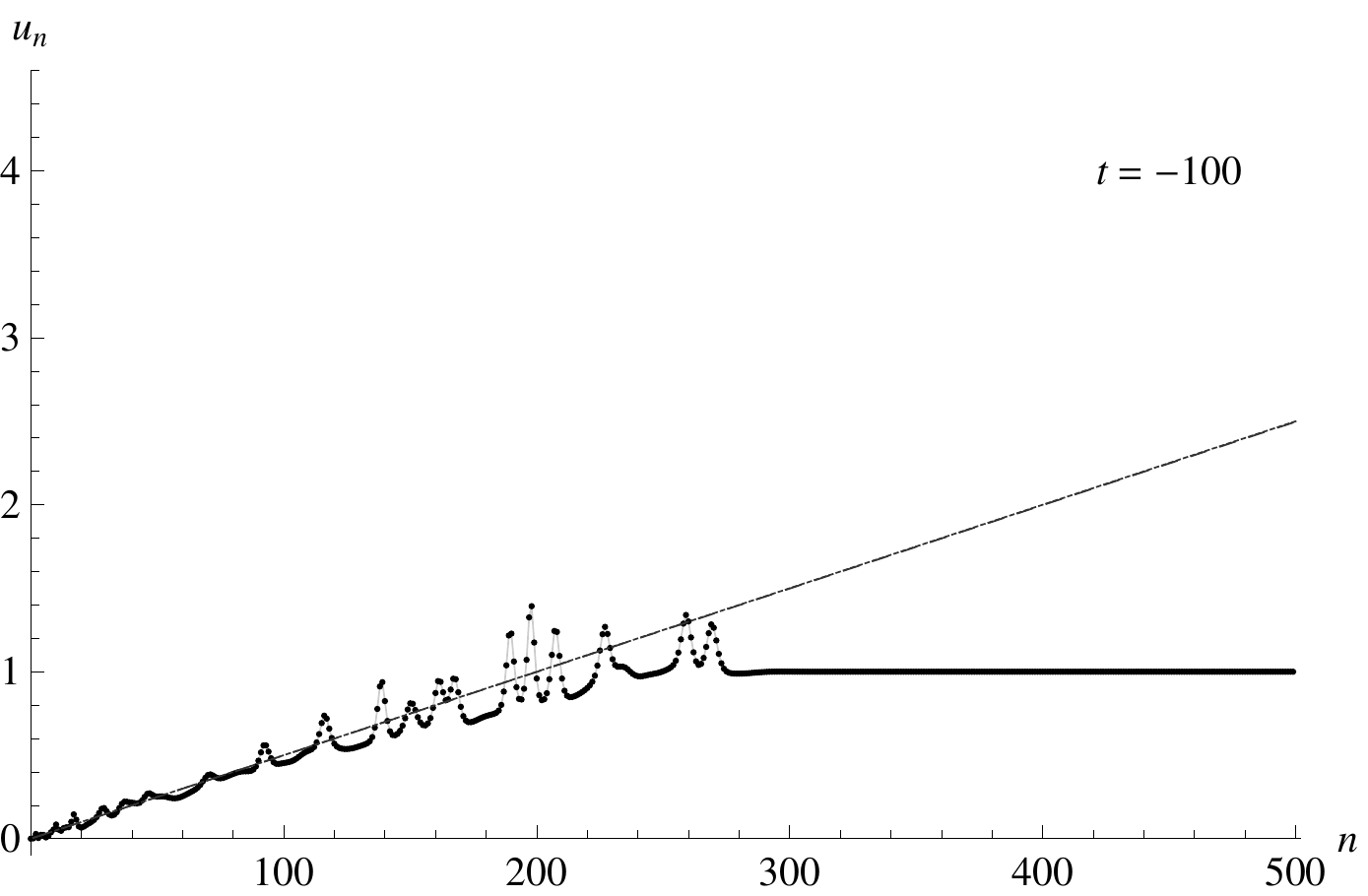}\qquad\includegraphics[width=70mm]{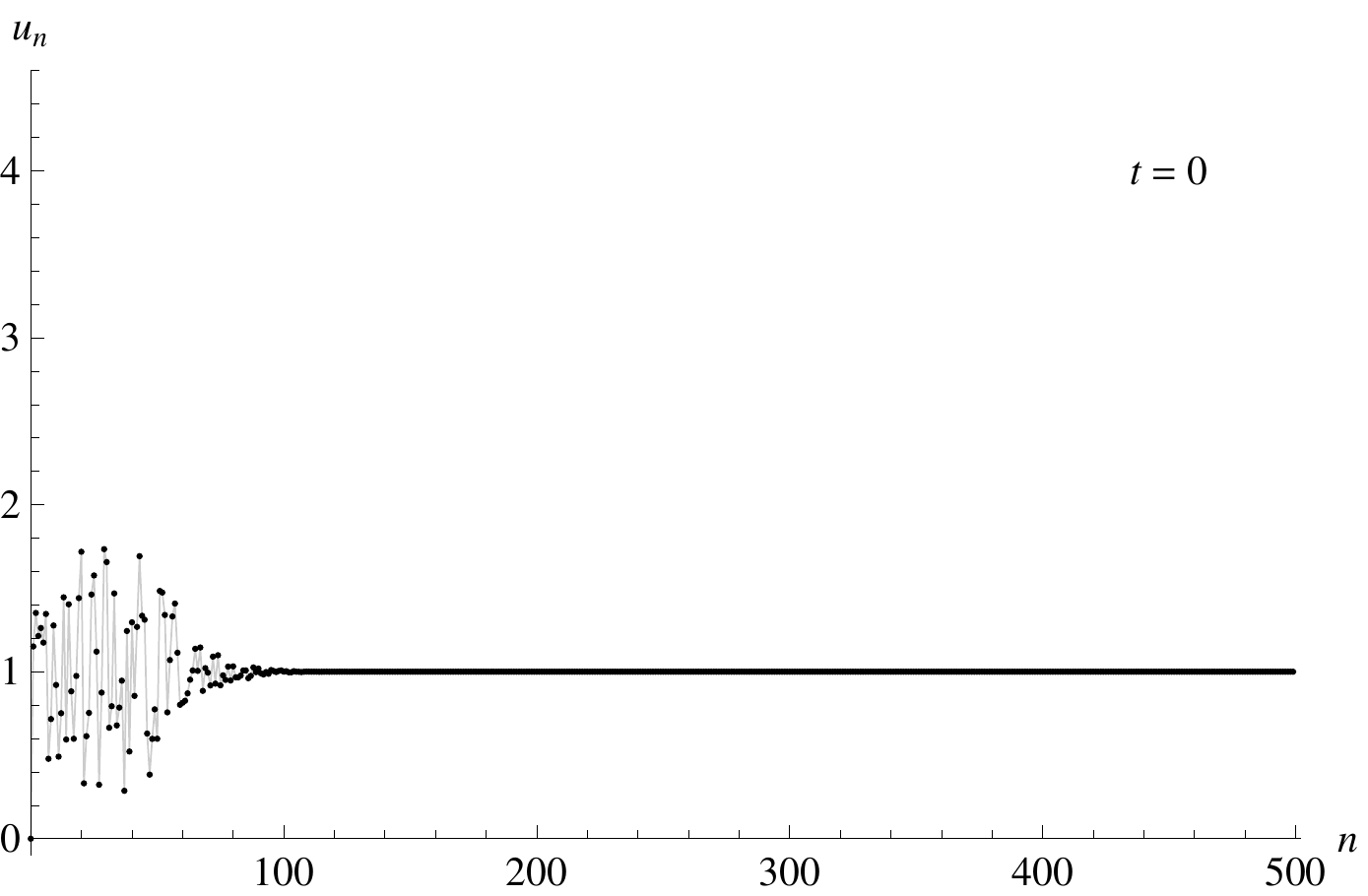}}
\centerline{\includegraphics[width=70mm]{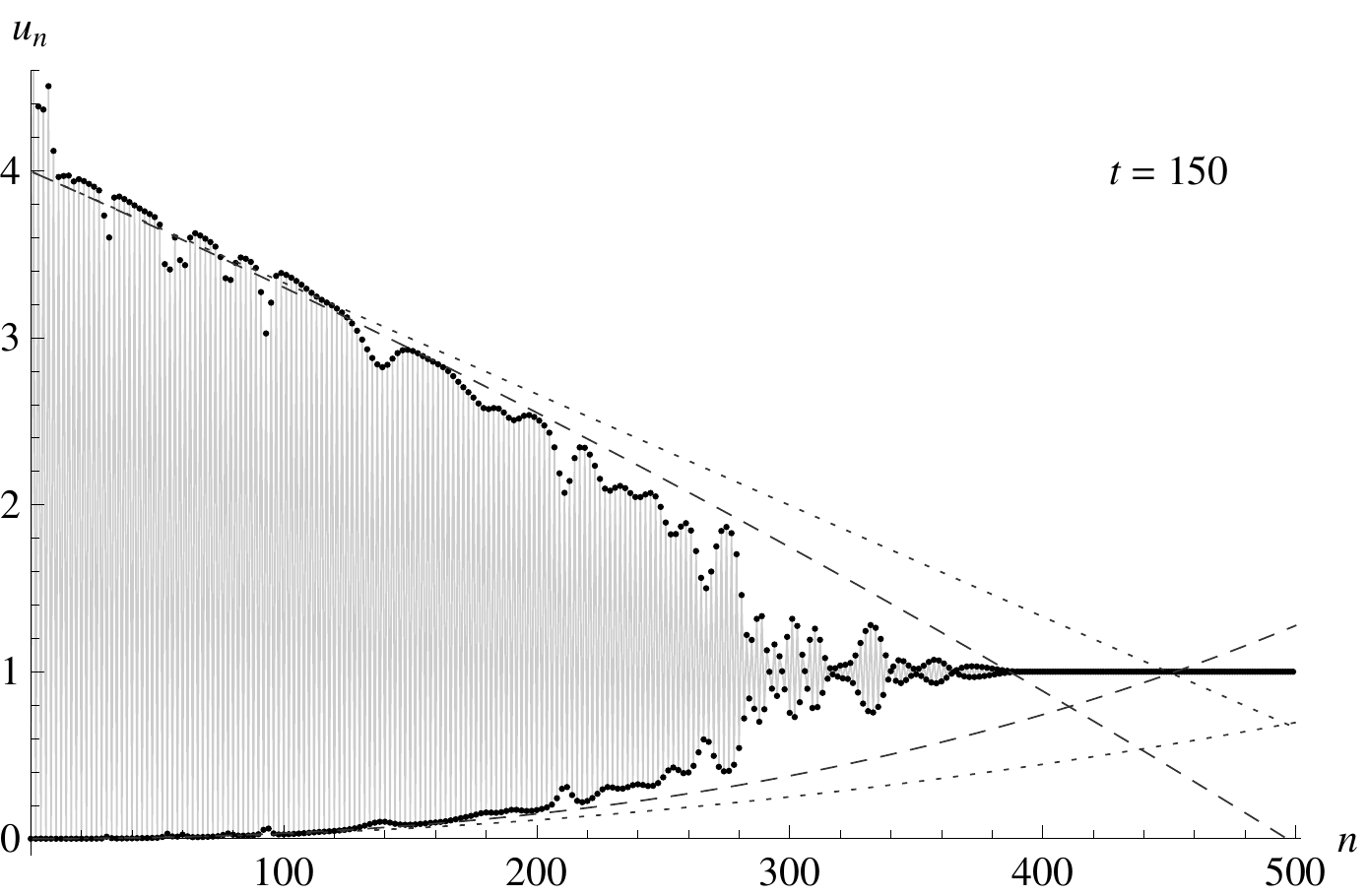}}
\caption{Solution with random perturbation of the initial unit step.}
\label{fig:rand}
\end{figure}

The initial data are changed for other admissible values of $b$ and $d$, but the general behaviour of the solution remains the same, exactly as for the case of solutions described in Section \ref{s:regular}. Moreover, the numeric experiments demonstrate that this mode is stable with respect to small enough perturbations of the step-like initial data (of course, under the condition that the boundary value $u_0=0$ is not changed). Fig.\,\ref{fig:rand} shows the evolution of the initial data
\[
 u_n(0)=1+0.9\exp(-0.001(n-30)^2)r_n,\quad n>0,
\] 
where $r_n$ is a random value uniformly distributed in $[-1,1]$. This perturbation leads to appearance of soliton-like structures on the pure decay solution background, but the overall asymptotics does not change. 

To determine the asymptotics, we use the formal expansion of the solution of the Riccati equation (\ref{u1'}) 
\[
 u_1=q_0+q_1t^{-1}+q_2t^{-2}+\dots 
\]
Substitution into equation proves that the leading term may take the values $q_0=4$ or $q_0=0$, while the subsequent coefficients are computed uniquely. This gives two series which correspond to the different asymptotics at $t\to\pm\infty$, which is easy to see by comparing with known asymptotic formulas \cite{AS}: 
\[
 M(\alpha,\beta,z)=\left\{
 \begin{aligned}
  &\frac{\Gamma(\beta)}{\Gamma(\alpha)}e^zz^{\alpha-\beta}\bigl(1+O(|z|^{-1})\bigr),  & \mathop{\rm Re} z>0,\\ 
  &\frac{\Gamma(\beta)}{\Gamma(\beta-\alpha)}(-z)^{-\alpha}\bigl(1+O(|z|^{-1})\bigr), & \mathop{\rm Re} z<0. 
 \end{aligned}\right.
\]
From here it follows, for the function $u_1=f'/f$ and for real $z=4t$, that $u_1\sim 4-(\alpha-\beta)t^{-1}+\dots$ at $t\to+\infty$ and $u_1\sim 0+\alpha t^{-1}+\dots$ at $t\to-\infty$. Next, the expansions for all $u_n$ are obtained by intermediate use of equation (\ref{u1'}) and the lattice equation (\ref{ut}). We wil restrict ourselves by two first terms of the asymptotics.

\begin{statement}\label{st:uinf}
Consider the solution of the lattice (\ref{ut}) with $u_0=0$ and the initial data $u_n(0)>0$ for $n>0$ defined by equations (\ref{und00}), (\ref{bdnek}) with $a=1$. Then the following asymptotic formulas are valid: for $t\to+\infty$
\[
 u_n=
  \left\{
  \begin{aligned}
   & \frac{n(n-2d+b)}{16t^2}+\frac{n(n-2d+b)(2n-2d+3b)}{128t^3}+O(t^{-4}),\quad n=0,2,\dots,\\
   & 4-\frac{2n-2d+b}{2t}\\
   & \qquad -\frac{n^2-(2d-3b)n+b^2-2bd+1}{16t^2}+O(t^{-3}),\quad n=1,3,\dots
  \end{aligned}
  \right.
\]
and for $t\to-\infty$
\[
 u_n=
  \left\{
  \begin{aligned}
   & -\frac{n}{2t}+\frac{(2d-b)n}{16t^2}+O(t^{-3}),\quad n=0,2,\dots,\\
   & -\frac{n-b}{2t}+\frac{(2d-b)(n-b)}{16t^2}+O(t^{-3}),\quad n=1,3,\dots
  \end{aligned}
  \right.
\]
\end{statement}
\begin{proof}
We find from equation (\ref{u1'}) that
\[
 u_1(t)=\left\{
  \begin{aligned} 
  & 4+\frac{2d-b-2}{2t}+\frac{(b+1)(2d-b-2)}{16t^2}+\dots,\quad t\to+\infty,\\
  & \frac{b-1}{2t}+\frac{(b-1)(b-2d)}{16t^2}+\dots,\quad t\to-\infty.
 \end{aligned}\right.
\]
It is easy to prove by induction that expansions for $u_n$ at $t\to+\infty$ are of the following form, depending on the parity of $n$:
\[
 u_{2j}=\frac{p_{j,2}}{t^2}+\frac{p_{j,3}}{t^3}+\dots,\quad
 u_{2j+1}=4+\frac{q_{j,1}}{t}+\frac{q_{j,2}}{t^2}+\dots
\]
and that the substitution into (\ref{ut}) gives the difference equations for the coefficients
\begin{gather*}
 q_{j,1}-q_{j-1,1}=-2,\quad 4p_{j+1,2}-4p_{j,2}=-q_{j,1},\\
 p_{j,2}(q_{j,2}-q_{j-1,2})=-p_{j,3},\quad 16p_{j+1,3}-16p_{j,3}+8q_{j,2}=q^2_{j,1}.
\end{gather*}
Similarly, the expansions at $t\to-\infty$ start from $t^{-1}$ for all $n$: 
\[
 u_{2j}=p_{j,1}t^{-1}+p_{j,2}t^{-2}+\dots,\quad u_{2j+1}=q_{j,1}t^{-1}+q_{j,2}t^{-2}+\dots
\] 
and the coefficients are governed by equations
\begin{gather*}
 q_{j,1}-q_{j-1,1}=-1,\quad p_{j+1,1}-p_{j,1}=-1,\\ 
 p_{j,1}(q_{j,2}-q_{j-1,2})=-p_{j,2},\quad  q_{j,1}(p_{j+1,2}-p_{j,2})=-q_{j,2}.
\end{gather*}
The initial data for these equations are given by the coefficients $q_{0,1},q_{0,2}$ of the above series for $u_1$ and the values $p_{0,k}=0$ corresponding to $u_0=0$. In both cases, the solution is easily obtained as polynomials in $j$ 
and we obtain the required formulas by returning to the variable $n$.
\end{proof}

For $t>0$, one can obtain a rough estimate of the decay zone by constructing a triangular region bounded by the plots of one or two terms of the asymptotic expansions, as shown on figs.\;\ref{fig:t25} and \ref{fig:rand}. In particular, an upper bound for the wedge point $n_0(t)$ of the decay zone can be obtained by solving the inequality $4-\frac{2n_0-2d+b}{2t}>1$, which gives, apparently, $n_0<3t+\const$. More accurate estimates can be obtained by taking the next asymptotic terms. For the negative $t$, the solution is well approximated by the first term of the asymptotics.

We conclude this section with a note on the conservation laws of the Volterra lattice, that is, relations of the form
\begin{equation}\label{conserv}
 \frac{d}{dt}\rho^{(k)}_n=\sigma^{(k)}_{n+1}-\sigma^{(k)}_n,
\end{equation}
where $\rho^{(k)}_n$ and $\sigma^{(k)}_n$ depend on a finite number of variables $u_n$. Three simplest conservation laws are given by
\begin{gather*}
 \rho^{(0)}_n=\log u_n,~~ \sigma^{(0)}_n=u_{n-1}+u_n,\qquad
 \rho^{(1)}_n=u_n,~~ \sigma^{(1)}_n=u_{n-1}u_n,\\
 \rho^{(2)}_n=\frac{1}{2}u^2_n+u_nu_{n+1},~~ \sigma^{(2)}_n=u_{n-1}u_n(u_n+u_{n+1}).
\end{gather*}
In the case of the problem on the whole line and with the initial data which have the same constant asymptotics for $n\to\pm\infty$, it follows from (\ref{conserv}) that the quantities $H_k=\sum_n(\rho^{(k)}_n-r^{(k)})$ are preserved, where the summation is taken over all integer $n$ and the constant $r^{(k)}$ is chosen so that the sum is well defined. For the lattice truncated by $u_0=0$, the analogous sums over $n>0$ do not preserve, since
\[
 \frac{d}{dt}H_k=\lim_{n\to\infty}\sigma^{(k)}_n-\sigma^{(k)}_1\ne0.
\]
For solutions with the asymptotics $u_n\to1$ for $n\to\infty$, the sums are regularized as follows
\[
 H_0=\sum^\infty_{n=1}\log u_n,\quad
 H_1=\sum^\infty_{n=1}(u_n-1),\quad
 H_2=\sum^\infty_{n=1}\Bigl(\frac{1}{2}u^2_n+u_nu_{n+1}-\frac{3}{2}\Bigr),
\]
and we have $\sigma^{(0)}_\infty=2$, $\sigma^{(0)}_1=u_1$;
$\sigma^{(1)}_\infty=1$, $\sigma^{(1)}_1=0$ and 
$\sigma^{(2)}_\infty=2$, $\sigma^{(2)}_1=0$. Then
\[
 \frac{d}{dt}H_0=2-u_1,\quad \frac{d}{dt}H_1=1,\quad \frac{d}{dt}H_2=2,
\]
and since all three sums are equal to 0 at $t=0$, hence
\[
 H_0= \int^t_0(2-u_1(\tau))d\tau,\quad H_1=t,\quad H_2=2t.
\]

%-------------------------------------------------------------------------------
\section{Determinant identities}\label{s:det}

In addition to the recurrent relations (\ref{recD}), (\ref{rec0}), there exists the Wronskian representation of the Volterra lattice solution on the half-line, which goes back to the Leznov paper \cite{Leznov_1980}. It is not very convenient for a practical computing of solutions, but we will show that using it together with explicit expressions for $u_1(t)$ and $u_n(0)$ makes possible to get nontrivial identities for some number sequences.

\begin{statement}\label{st:w}
The solution of the lattice (\ref{ut}) on the half-line $n\ge0$, such that $u_0=0$ and $u_1=f'/f$, with an arbitrary infinitely differentiable function $f(t)$, is of the form
\begin{equation}\label{uw}
 u_n=\frac{w_{n-3}w_n}{w_{n-2}w_{n-1}},\quad n=0,1,2,\dots,
\end{equation}
where $w_{-3}=0$, $w_{-2}=w_{-1}=1$ and, for $k\ge0$,  
\begin{equation}\label{ww}
 w_{2k}=
 \left|\begin{matrix}
   f       & f'        & \dots  & f^{(k)}  \\
   f'      & f''       & \dots  & f^{(k+1)}\\
   \vdots  & \vdots    & \ddots &\vdots    \\
   f^{(k)} & f^{(k+1)} & \dots  & f^{(2k)}
 \end{matrix}\right|,\quad
 w_{2k+1}=
 \left|\begin{matrix}
   f'       & f''       & \dots  & f^{(k+1)}  \\
   f''      & f'''      & \dots  & f^{(k+2)}\\
   \vdots   & \vdots    & \ddots &\vdots    \\
   f^{(k+1)}& f^{(k+2)} & \dots  & f^{(2k+1)}
 \end{matrix}\right|.
\end{equation}
\end{statement}
\begin{proof}
We will prove that $w_n$ satisfy equations
\begin{equation}\label{wt}
 w_nw'_{n+1}-w'_nw_{n+1}=w_{n-1}w_{n+2},\quad n=-2,-1,0,1,\dots,
\end{equation}
then it is easy to check that substitution (\ref{uw}) gives a (unique) solution of the lattice equations (\ref{ut}) with $u_0=0$ and $u_1=f'/f$. 

For $n=-2,-1$, the relations (\ref{wt}) are verified directly. For $n\ge0$, let $W(A)$ denote the Wronskian of an arbitrary finite sequence $A$ of smooth functions (possibly empty). For $n=2k$ we set $A=f^{(1)},\dots,f^{(k)}$, then
\begin{alignat*}{2}
 &w_{n-1}=W(A), &\quad&     w_n=W(f^{(0)},A) =(-1)^kW(A,f^{(0)}),\\
 &w_{n+1}=W(A,f^{(k+1)}),&& w_{n+2}=W(f^{(0)},A,f^{(k+1)}) =(-1)^kW(A,f^{(0)},f^{(k+1)}).
\end{alignat*}
Similarly, for $n=2k+1$ we set $A=f^{(0)},\dots,f^{(k)}$, then
\begin{alignat*}{2}
 &w_{n-1}=W(A), &\quad&     w_n=W(1,A)=(-1)^{k+1}W(A,1),\\
 &w_{n+1}=W(A,f^{(k+1)}),&& w_{n+2}=W(1,A,f^{(k+1)})=(-1)^{k+1}W(A,1,f^{(k+1)}).
\end{alignat*}
In both cases, equation (\ref{wt}) is satisfied due to the identity
\[
 W(A,b)\frac{d}{dt}W(A,c)-\frac{d}{dt}W(A,b)W(A,c)=W(A)W(A,b,c),
\]
where $b,c$ are arbitrary smooth functions. In order to prove it, it is sufficient to consider both left and right hand sides as the differential operators with respect to $c$ and to compare their kernels and the coefficients at the highest derivative.
\end{proof}

The determinants of the form (\ref{ww}) for the number sequences $f_0,f_1,f_2,\dots$ are actively studied in combinatorics. Recall that the Hankel transformation for such a sequence is the sequence of determinants of size $n\times n$ with $(i,j)$-th element equal to $f_{i+j-2}$. This mapping is not one-to-one, since the determinant of size $n$ involves $2n-1$ members of the sequence. To get a one-to-one mapping, one can use simultaneously Hankel transforms for the sequence itself and the sequence without the zero member, provided that all determinants do not vanish.

For instance, the Hankel transform of the Catalan numbers $1$, $1$, $2$, $5$, $14$, $42$, $132$, $429,\dots$ is the sequence $1,1,1,\dots$ and the same is true for the sequence of the Catalan numbers with the first member dropped \cite{Aigner_1999, Stanley_1999, Layman_2001}. In \cite{Adler_Shabat_2018}, we associated these identities of the Catalan numbers with the Wronskian representation of the solution with the initial data $u_0=0$, $u_n(0)=1$, $n>0$ and derived the Riccati equation for $u_1(t)$ from this.

Now we are able to reverse and to generalize this result. Indeed, the Wronskians $w_{2k}$ and $w_{2k+1}$ define the Hankel transformations for the sequences $f,f',\dots$ and $f',f'',\dots$, respectively. By setting $t=0$ in (\ref{ww}), we obtain the Hankel transformations for the coefficients of the Taylor expansion of the function $f(t)$ (here, the regularity for other values of $t$ is not necessary) and it remains to compare the result with the known initial data by substituting into equation (\ref{uw}) at $t=0$.

\begin{statement}\label{st:Hankel}
Let a solution of the lattice equation (\ref{ut}) on the half-line $n\ge0$ be given by equations (\ref{uw}), (\ref{ww}) with the function $f$ given by the Taylor expansion
\[
 f(t)=f_0+f_1t+\cdots+f_n\frac{t^n}{n!}+\cdots,
\]
then, for $k=0,1,2,\dots$, 
\begin{equation}\label{wwu0}
\begin{gathered}
 h_{2k}=\left|\begin{matrix}
   f_0    & \dots  & f_k     \\
   \vdots & \ddots &\vdots   \\
   f_k    & \dots  & f_{2k}
 \end{matrix}\right|  =  \prod^k_{j=1}(u_{2j-1}(0)u_{2j}(0))^{k+1-j},\qquad\\
 h_{2k+1}=\left|\begin{matrix}
   f_1     & \dots  & f_{k+1}  \\
   \vdots  & \ddots &\vdots    \\
   f_{k+1} & \dots  & f_{2k+1}
 \end{matrix}\right|  =  u^{k+1}_1(0)\prod^k_{j=1}(u_{2j}(0)u_{2j+1}(0))^{k+1-j}.
\end{gathered}
\end{equation}
\end{statement}
\begin{proof}
By setting $h_n=w_n(0)$, we obtain $h_{-2}=h_{-1}=h_0=1$ and the recurrent relation
\[
 h_nh_{n-3}=u_n(0)h_{n-1}h_{n-2},\quad n=1,2,\dots, 
\]
which proves the statement by induction.
\end{proof}

We note that some nontrivial identities follows from here even for the explicit solution
\[
 u_{2k-1}=e^t,\quad u_{2k}=k
\]
which is a particular case of solution (\ref{expsol}) at $\alpha=0$, $\beta=-1/2$, $\gamma=-2$ and $\delta=0$. In this case we find, by solving equation $f'/f=u_1=e^t$, that
\[
 f(t)=e^{e^t-1}= \sum^\infty_{n=0}B_n\frac{t^n}{n!},
\]
where $B_n$ are the Bell numbers $1,1,2,5,15,52,203,\dots$ enumerating the partitions of a set of $n$ elements. Here $u_{2k-1}(0)=1$, $u_{2k}(0)=k$ and one obtains easily that $h_{2k}=h_{2k+1}=1!\cdots k!$ (the superfactorial of $k$). This result is known since 1978~\cite{Ehrenborg_2000}.

Now, let us apply Statement \ref{st:Hankel} to solutions desribed in the previous section (however, now we do not set $a=1$). Due to the known expansion of the Kummer function, we have
\begin{equation}\label{Kummer}
 f(t)= M(\alpha,\beta,4at)
     = 1+4a\frac{\alpha}{\beta}t+(4a)^2\frac{(\alpha)_2}{(\beta)_2}\frac{t^2}{2!}+\dots
        +(4a)^n\frac{(\alpha)_n}{(\beta)_n}\frac{t^n}{n!}+\dots,
\end{equation}
where $(\alpha)_n$ is the Pochhammer symbol 
\[
 (\alpha)_n=\alpha(\alpha+1)\cdots(\alpha+n-1),\quad (\alpha)_0=1. 
\]
In addition, let us denote
\[
 ((\alpha))_n= (\alpha)_1\cdots(\alpha)_n= \alpha^n(\alpha+1)^{n-1}\cdots(\alpha+n-1)^1,\quad ((\alpha))_0=1; 
\]
in particular, the superfactorial is denoted as $((1))_n=1!\cdots n!$.

\begin{statement}\label{st:Poch-Hankel}
Let
\begin{gather*}
 f_n=(4a)^n\frac{(\alpha)_n}{(\beta)_n},\quad n=0,1,2,\dots,\\
 \alpha\ne -n,\quad \beta\ne -n,\quad \alpha-\beta\ne n,\quad n=0,1,2,\dots,
\end{gather*}
then, for $k=0,1,2,\dots$, 
\begin{equation}\label{Poch-Hankel}
\begin{gathered}
 h_{2k}=\left|\begin{matrix}
   f_0    & \dots  & f_k     \\
   \vdots & \ddots &\vdots   \\
   f_k    & \dots  & f_{2k}
 \end{matrix}\right|  = \frac{((1))_k(4a)^{k(k+1)}((\alpha))_k((\beta-\alpha))_k}{(\beta)^{k+1}_k((\beta+k))_k},\\
 h_{2k+1}=\left|\begin{matrix}
   f_1     & \dots  & f_{k+1}  \\
   \vdots  & \ddots &\vdots    \\
   f_{k+1} & \dots  & f_{2k+1}
 \end{matrix}\right|  = \frac{((1))_k(4a)^{(k+1)^2}((\alpha))_{k+1}((\beta-\alpha))_k}{(\beta)^{k+1}_k((\beta+k))_{k+1}},
\end{gathered}
\end{equation}
\end{statement}
\begin{proof}
Set 
\[
 b=a(1-2\alpha),\quad d=\frac{3}{2}-\beta,
\]    
then the initial data (\ref{und00}) take the form
\[
 \begin{aligned}
  u_1(0)=4a\frac{\alpha}{\beta},\quad
   &u_{2k}(0)=4a\frac{k(\beta-\alpha+k-1)}{(\beta+2k-2)(\beta+2k-1)},\\
   &u_{2k+1}(0)=4a\frac{(\alpha+k)(\beta+k-1)}{(\beta+2k-1)(\beta+2k)},
 \end{aligned}\quad k=1,2,3,\dots,
\]
and the inequalities for $\alpha$ and $\beta$ coincide exactly with the conditions (\ref{bdnek}) which guarantee that all numerators and denominators are not 0. According to Statements \ref{st:Riccati} and \ref{st:w}, the corresponding regular (at $t=0$) solution of the Volterra lattice is defined by equations (\ref{uw}), (\ref{ww}) with function (\ref{Kummer}) and we only have to substitute the initial data into (\ref{wwu0}) and to arrange the factors. 
\end{proof}

In particular, the example with the Catalan numbers $f_n=\dfrac{(2n)!}{(n+1)!n!}$ corresponds to the choice $a=1$, $b=0$ and $d=-\frac{1}{2}$ (or, $\alpha=\frac{1}{2}$ and $\beta=2$). In this case the initial data are $u_n(0)=1$ for $n>0$ 
and instead of the general formula, it is easier to use the recurrence relation $h_nh_{n-3}=u_n(0)h_{n-1}h_{n-2}$ directly, which immediately gives that all $h_n=1$.

Similarly, the central binomial coefficients $f_n=\dfrac{(2n)!}{(n!)^2}$ correspond to the choice $a=1$, $b=0$ and $d=\frac{1}{2}$ (or, $\alpha=\frac{1}{2}$ and $\beta=1$). In this case $u_1(0)=2$ and $u_n(0)=1$ for $n>1$ and the recurrent relation yields $h_{2k}=2^k$, $h_{2k+1}=2^{k+1}$. This example is also known in the combinatorics.

%-------------------------------------------------------------------------------
\subsubsection*{Acknowledgements}

This work was carried out under the State Assignment 0033-2019-0006 (Integrable systems of mathematical physics) of the Ministry of Science and Higher Education of the Russian Federation.

%-------------------------------------------------------------------------------

\end{document}